\documentstyle[12pt]{article}

\catcode`\@=11
\long\def\@makefntext#1{
\protect\noindent \hbox to 3.2pt {\hskip-.9pt
$^{{\ninerm\@thefnmark}}$\hfil}#1\hfill}		%CAN BE USED

\def\@makefnmark{\hbox to 0pt{$^{\@thefnmark}$\hss}}  %ORIGINAL

\def\ps@myheadings{\let\@mkboth\@gobbletwo
\def\@oddhead{\hbox{}
\rightmark\hfil\ninerm\thepage}
\def\@oddfoot{}\def\@evenhead{\ninerm\thepage\hfil
\leftmark\hbox{}}\def\@evenfoot{}
\def\sectionmark##1{}\def\subsectionmark##1{}}

\renewcommand{\thefootnote}{\fnsymbol{footnote}}

\newcounter{sectionc}\newcounter{subsectionc}\newcounter{subsubsectionc}
\renewcommand{\section}[1] {\vspace*{0.6cm}\addtocounter{sectionc}{1}
\setcounter{subsectionc}{0}\setcounter{subsubsectionc}{0}\noindent
	{\normalsize\bf\thesectionc. #1}\par\vspace*{0.4cm}}
\renewcommand{\subsection}[1] {\vspace*{0.6cm}\addtocounter{subsectionc}{1}
	\setcounter{subsubsectionc}{0}\noindent
	{\normalsize\it\thesectionc.\thesubsectionc. #1}\par\vspace*{0.4cm}}
\renewcommand{\subsubsection}[1]
{\vspace*{0.6cm}\addtocounter{subsubsectionc}{1}
	\noindent {\normalsize\rm\thesectionc.\thesubsectionc.\thesubsubsectionc.
	#1}\par\vspace*{0.4cm}}

\newcounter{appendixc}
\newcounter{subappendixc}[appendixc]
\newcounter{subsubappendixc}[subappendixc]

\renewcommand{\appendix}[1] {\vspace*{0.6cm}
        \refstepcounter{appendixc}
        \setcounter{figure}{0}
        \setcounter{table}{0}
        \setcounter{equation}{0}
        \renewcommand{\thefigure}{\Alph{appendixc}.\arabic{figure}}
        \renewcommand{\thetable}{\Alph{appendixc}.\arabic{table}}
        \renewcommand{\theappendixc}{\Alph{appendixc}}
        \renewcommand{\theequation}{\Alph{appendixc}.\arabic{equation}}
        \noindent{\bf Appendix \theappendixc #1}\par\vspace*{0.4cm}}

\def\abstracts#1{{

\centering{\begin{minipage}{12.2truecm}\footnotesize\baselineskip=12pt\noindent
	\centerline{\footnotesize ABSTRACT}\vspace*{0.3cm}
	\parindent=0pt #1
	\end{minipage}}\par}}

\renewenvironment{thebibliography}[1]
	{\begin{list}{\arabic{enumi}.}
	{\usecounter{enumi}\setlength{\parsep}{0pt}
\setlength{\leftmargin 1.25cm}{\rightmargin 0pt}
	 \setlength{\itemsep}{0pt} \settowidth
	{\labelwidth}{#1.}\sloppy}}{\end{list}}

\newcounter{itemlistc}
\newcounter{romanlistc}
\newcounter{alphlistc}
\newcounter{arabiclistc}

\newcommand{\fcaption}[1]{
        \refstepcounter{figure}
        \setbox\@tempboxa = \hbox{\footnotesize Fig.~\thefigure. #1}
        \ifdim \wd\@tempboxa > 6in
           {\begin{center}
        \parbox{6in}{\footnotesize\baselineskip=12pt Fig.~\thefigure. #1}
            \end{center}}
        \else
             {\begin{center}
             {\footnotesize Fig.~\thefigure. #1}
              \end{center}}
        \fi}

\newcommand{\tcaption}[1]{
        \refstepcounter{table}
        \setbox\@tempboxa = \hbox{\footnotesize Table~\thetable. #1}
        \ifdim \wd\@tempboxa > 6in
           {\begin{center}
        \parbox{6in}{\footnotesize\baselineskip=12pt Table~\thetable. #1}
            \end{center}}
        \else
             {\begin{center}
             {\footnotesize Table~\thetable. #1}
              \end{center}}
        \fi}

\def\@citex[#1]#2{\if@filesw\immediate\write\@auxout
	{\string\citation{#2}}\fi
\def\@citea{}\@cite{\@for\@citeb:=#2\do
	{\@citea\def\@citea{,}\@ifundefined
	{b@\@citeb}{{\bf ?}\@warning
	{Citation `\@citeb' on page \thepage \space undefined}}
	{\csname b@\@citeb\endcsname}}}{#1}}

\newif\if@cghi
\def\cite{\@cghitrue\@ifnextchar [{\@tempswatrue
	\@citex}{\@tempswafalse\@citex[]}}
\def\citelow{\@cghifalse\@ifnextchar [{\@tempswatrue
	\@citex}{\@tempswafalse\@citex[]}}
\def\@cite#1#2{{$\null^{#1}$\if@tempswa\typeout
	{IJCGA warning: optional citation argument
	ignored: `#2'} \fi}}

 1
 1
 1

\font\ninerm=cmr9

\begin{document}

\rightline{Preprint Number:
\parbox[t]{35mm}{ADP-95-28/T182 \\
		 hep-ph/9505208}}
\vskip12pt
\rightline{\it Invited talk given at the International RCNP Workshop on Color}
\rightline{\it Confinement and Hadrons, Osaka, Japan, March 22-24, 1995.}
\rightline{\it (To appear in the conference proceedings).}

\vspace{0.6cm}

\centerline{\normalsize\bf SOLVING THE BETHE--SALPETER EQUATION IN}
\baselineskip=22pt
\centerline{\normalsize\bf MINKOWSKI SPACE: SCALAR THEORIES}
\baselineskip=16pt

%\vfill
\vspace*{0.6cm}
\centerline{\footnotesize Kensuke Kusaka$^a$ and
\underline{Anthony G. Williams}$^b$}
\baselineskip=13pt
\centerline{\footnotesize\it Department of Physics and Mathematical Physics,}
\centerline{\footnotesize\it University of Adelaide}
\baselineskip=12pt
\centerline{\footnotesize\it South Australia 5005, Australia}
\centerline{\footnotesize $^a$E-mail: kkusaka@physics.adelaide.edu.au}
\centerline{\footnotesize $^b$E-mail: awilliam@physics.adelaide.edu.au}
%\vspace*{0.3cm}
%\centerline{\footnotesize and}
%\vspace*{0.3cm}
%\centerline{\footnotesize Kensuke Kusaka}
%\baselineskip=13pt
%\centerline{\footnotesize\it Group, Company, Address, City, State ZIP/Zone,
%Country}

%\vfill
\vspace*{0.9cm}
\abstracts{
The Bethe-Salpeter (BS) equation for scalar-scalar bound
states in scalar theories without derivative coupling
is formulated and solved in Minkowski space.
This is achieved using the perturbation theory integral representation
(PTIR), which allows these amplitudes to be
expressed as integrals over weight functions and known singularity
structures and hence allows us to convert the BS equation
into an integral equation involving weight functions.
We obtain numerical solutions using this formalism
for a number of scattering kernels to illustrate the generality of the
approach.  It applies even when the na\"{\i}ve
Wick rotation is invalid. As a check we verify, for example, that
this method applied to the special case of the massive ladder exchange
kernel reproduces the same results as are obtained by Wick rotation.
}

%\vspace*{0.6cm}
\normalsize\baselineskip=15pt
\setcounter{footnote}{0}
\renewcommand{\thefootnote}{\alph{footnote}}
\section{Introduction}
There has been considerable recent interest in
covariant descriptions
of bound states, for example, in conjunction with model calculations
of high-energy processes such as deep inelastic scattering.
A fully covariant description of composite
bound states is essential for the understanding of hadronic structure
over the full range of available momentum transfer.
In a relativistic field theory
the two-body component of a bound state is described
by the appropriate proper (i.e., one-particle irreducible)
three-point vertex function or, equivalently, by the Bethe-Salpeter (BS)
amplitude\cite{Nakanishi_survey}.
The BS amplitude satisfies an integral equation
whose kernel has singularities due to the Minkowski metric.
The resultant solutions are not functions but mathematical distributions.
The singularity structure of these distributions
makes it difficult to handle the BS equation
numerically in Minkowski space.

One approach to dealing with the difficulties presented by the Minkowski-space
metric is to perform an analytic continuation in the
relative-energy variable $p^0$, which is the so-called ``Wick rotation''
\cite{Wick}.  This has been widely used as a means of solving model
BS equations, e.g., see Ref.~3 and references therein.
The special case of the ladder BS equation is solved as a function
of Euclidean relative momentum in the standard treatment.
However it is necessary to rotate Euclidean solutions back to
Minkowsi ones in order to apply the BS amplitude (vertex function) to
calculations of absorptive part of scattering amplitudes for example.
The difficulties associated with this approach in the general case
arise from the fact that
since the total four-momentum $P$ must remain timelike then
the kernel becomes complex, so that one needs to perform analytic
continuation numerically to obtain Minkowski solutions.
In addition, when one uses a ``dressed'' propagator
for the constituent particles or more complicated kernels
in the BS equation, the validity of the Wick rotation becomes
highly  nontrivial.  For example, essentially all of
the dressed propagators studied previously in Wick-rotated
Dyson-Schwinger equation
approach contain complex ``ghost'' poles\cite{TheReview}.
Hence, the na\"{\i}ve Wick rotation obtained by the simple
transcription of Minkowski metric to the Euclidean metric
and vice versa is not valid in general.
So while the Euclidean-based approaches certainly play a very important role
and can be useful in model calculations, it is
preferable to formulate and solve the BS equation directly in
Minkowski space.
Here we present a method to solve the BS equation without Wick
rotation by making use of the perturbation theory integral representation
(PTIR) for the BS amplitude\cite{Nakanishi_graph}.

The PTIR is a natural extension of the spectral representation for
a two-point Green's function to
an $n$-point function in a relativistic field theory.  Since
the Feynman parametric integral always exists for any Feynman diagram
in perturbative calculations,
one can always define the integral representation
such that the number of independent integration parameters is equal
to that of invariant squares of external momenta.
Each Feynman diagram contributes to the
weight distribution of the parametric integral, so that
the complete weight function for the renormalized $n$-point function
is identical to the infinite sum of Feynman diagrams for the renormalized
Lagrangian of the theory.  Hence, we see that
the PTIR of a particular renormalized $n$-point function is an integral
representation of the corresponding infinite sum of Feynman
diagrams for the renormalized theory with $n$ fixed external lines.

The PTIR method was first used in conjunction with a Wick rotation
by Wick and Cutkosky for solving a scalar-scalar bound state
with a massless scalar exchange
in the ladder approximation\cite{Wick,Cutkosky}.
This is now commonly referred to as the Wick-Cutkosky model.
They solved the BS equation in terms
of a single variable integral representation,
which is a special case of the PTIR.  After initial efforts to
solve scalar-scalar BSE's using the PTIR method\cite{Nakanishi_survey},
Nakanishi made a detailed and systematic study of the
PTIR\cite{Nakanishi_graph}.  Here we extend the PTIR method
to solve the BS equation with a broad class of scattering kernels,
by transforming the BSE in momentum space to an integral equation for
the weight function without relying on the Wick rotation.

\section{Scalar-Scalar BS Equation}
Let us consider a bound state of two spinless $\phi$ particles
having a mass $m$.
They interact with
each other through the exchange of other spinless particles.
The Bethe-Salpeter amplitude $\Phi(p,P)$ for the bound state having the
total momentum $P$ and the relative momentum $p$
obeys the following equation:
\begin{equation}
	[D(P/2+p)D(P/2-p)]^{-1}\Phi(p,P) = \int{d^4q\over (2\pi)^4i}
        I(p,q;P)\Phi(q,P)
	\label{BSE}
\end{equation}
where $D(q)$ is the propagator of $\phi$-particle.  We approximate it
with the free one: \( D^0(q)=1/(m^2-q^2-i\epsilon) \).
The scattering kernel $I(p,q;P)$ describes the process
$\phi\phi \rightarrow \phi\phi$, where
$p$ and $q$ are the initial and final relative momenta
respectively.  It is given by the infinite series of Feynman diagrams
which are two-particle irreducible
with respect to the initial and final pairs of constituent $\phi$
particles.
For purely scalar theories without derivative coupling we have
the formal expression for the full
renormalized scattering kernel
\begin{eqnarray}
    I(p,q;P) &=&
    \int\limits_{0}^{\infty } d\gamma\int\limits_{\Omega } d\vec\xi
     \left\{ \frac{ \rho_{st} (\gamma,\vec\xi) }
    { \gamma - \left[ \sum_{i=1}^{4}\xi_iq_i^2+\xi_5s+\xi_6t
     \right]-i\epsilon }
    \right.     \\
    &  &
    + \frac{ \rho_{tu}(\gamma,\vec\xi) }
    { \gamma - \left[ \sum_{i=1}^{4}\xi_iq_i^2+\xi_5t+\xi_6u \right]
     -i\epsilon }
     \left. + \frac{\rho_{us}(\gamma,\vec\xi) }
    {\gamma-\left[\sum_{i=1}^{4}\xi_iq_i^2+\xi_5u+\xi_6s\right]-i\epsilon}
    \right\}\;,\nonumber
    \label{krnl_PTIR}
\end{eqnarray}
where $q_i^2$ is the 4-momentum squared carried by $\phi_i$
and $s$,$t$ and $u$ are the usual Mandelstam variables.
This expression has been derived by Nakanishi (PTIR)\cite{Nakanishi_graph}.
Since only six of these are independent due to the relation
$q_1^2+q_2^2+q_3^2+q_4^2=s+t+u$, this is also the number of independent
$\xi$-parameters.  Hence, we need only introduce
the ``mass'' parameter $\gamma$ and
six dimensionless Feynman parameters $\xi_i$ with a constraint.
The symbol $\Omega$ denotes the integral region of
$\xi_i$ such that $\Omega\equiv\{\xi_i \,| \,0\leq\xi_i\leq 1, \,
\sum\xi_i=1 \, (i=1,\dots, 6)\}$.  The ``mass'' parameter $\gamma$
represents a spectrum of the scattering kernel.  The function
$\rho_{\rm ch}(\gamma,\vec\xi)$ gives the weight
of the spectrum arising from three different
channels which can be denoted ${\rm ch}= \{st\},\{tu\},\{us\}$.
This PTIR expression follows since any perturbative Feynman diagram for the
scattering kernel
can be written in this form, and hence this must also be true of their
sum.  In a perturbative calculation the weight function
$\rho_{\rm ch}$ is calculable as a power series of the coupling
constant for a given interaction Lagrangian.
%One can impose additional support properties
%on the kernel weight functions $\rho_{\rm ch}(\gamma,\vec\xi)$
%by analyzing general Feynman integrals for the theory of interest
%\cite{Nakanishi_graph}.  However, we do not impose any further conditions
%at this stage since we wish to incorporate more general cases such as
%separable kernels, which cannot
%be written as combinations of ordinary Feynman diagrams.

\section{PTIR for BS Amplitude}
\label{sec_PTIR_BS}

Let us now consider the PTIR of the BS amplitude itself.
Since the BS amplitude in the center-of-momentum
rest frame [i.e., with $P^\mu=(M,\vec 0)$] forms
an irreducible representation of the $O(3)$ rotation
group, we can label all of the
bound states with the usual 3-dimensional angular momentum quantum
numbers $\ell$ and $\ell_z$.
We can thus construct the integral representation
of the partial wave BS amplitude in the rest frame and the PTIR
automatically allows us to then boost to an arbitrary frame.

For simplicity, let us begin by considering the $s$-wave amplitude.
{}From general analysis of the PTIR for a one-particle-irreducible
3-point vertex function\cite{Nakanishi_graph},
one can write the BS amplitude for a
$s$-wave bound state
\begin{equation}
	\Phi(p,P)=-i\int\limits_{-\infty}^{\infty}d\alpha
		\int\limits_{-1}^{1}dz
		\frac{\varphi_n(\alpha,z)}
		{\left[m^2+\alpha-\left(p^2 + z p\cdot P+\frac{P^2}{4}\right)
			-i\epsilon\right]^{n+2}}\;,
	\label{BS_PTIR}
\end{equation}
where the ``mass'' parameter $\alpha$ represents a spectrum of the
BS amplitude and the weight function $\varphi_n(\alpha,z)$ gives
its weight.  One may regard the dependence of the weight function
on the dimensionless parameter $z$ as kinematical effects.
This two-variable PTIR is valid for the vertex function with
a timelike total momentum satisfying $0 < P^2 < 4m^2$.
We have introduced
a non-negative integer dummy parameter $n$.
We see that a partial integration of Eq.~(\ref{BS_PTIR}) with respect to
$\alpha$ connects weight functions with different $n$, i.e.,
\begin{equation}
	\varphi_{n+1}(\alpha,z)=(n+2)\int\limits_{-\infty}^\alpha
					d\alpha'\varphi_{n}(\alpha',z)\;,
	\label{rln_wghtfnc}
\end{equation}
provided that we have the boundary conditions
\begin{equation}
	\varphi_n(\alpha=-\infty,z)=0\; \quad {\rm and} \quad
	\lim_{\alpha\rightarrow\infty}
		\frac{\varphi_n(\alpha,z)}{\alpha^{n+1}}=0,
  \label{bc_PTIR}
\end{equation}
which must be satisfied in order for the expression (\ref{BS_PTIR})
to be meaningful.
While the non-negative integer $n$ is arbitrary,
a judicious choice can be advantageous
in numerical calculations, since the larger we take $n$,
the smoother the weight function becomes for a given $\Phi(p,P)$.

The extension of the PTIR to the BS amplitudes for higher partial waves, i.e,
those for bound states with non-zero angular momentum ($\ell>0$) is
straightforward.
The momentum-dependent structure, i.e., the denominator, of the
PTIR's for the bound state BS amplitude, is independent of
transformations under the little group belonging to the 3-momentum
$\vec p$ in the rest frame of the bound state, i.e., $P^\mu=(M,\vec 0)$.
Hence, for higher partial waves
with angular momentum quantum number
$\ell$ and third component $\ell_z$,
$\Phi$ can be written as the product
of the $\ell$-th order solid harmonic
${\cal Y}_\ell^{\ell_z}(\vec p)=|\vec p|^\ell Y^{\ell_z}_\ell(\vec p)$
and the corresponding PTIR for the scalar (i.e., s-wave) bound state
in this frame \cite{Nakanishi_survey}.
By considering the Lorentz boost from the rest frame to an arbitrary one
we have the integral representation of the partial wave BS amplitude
\begin{equation}
	\Phi^{[\ell,\ell_z]}(p,P)=-i{\cal Y}_\ell^{\ell_z}(\vec{p'})
		\int\limits_{-\infty}^{\infty}d\alpha
		\int\limits_{-1}^{1}dz
		\frac{\varphi^{[\ell]}_n(\alpha,z)}
		{\left[m^2+\alpha-\left(p^2 + z p\cdot P + P^2/4\right)
			-i\epsilon\right]^{n+2}}\;,
	\label{ptlw_BS_PTIR}
\end{equation}
where $P$ is an arbitrary timelike 4-vector with $P^2=M^2$
and $p'=\Lambda^{-1}(P)p$.  The Lorentz transformation
$\Lambda(P)$ connects $P$ and the bound-state rest frame 4-vector
$P'=(M,\vec 0)$, i.e, $P = \Lambda(P) P'$.
In the following sections
we will study the BS equation Eq.~(\ref{BSE}) in an arbitrary frame
in terms of this integral representation.

\section{BS Equation for the Weight Function}
\label{Integral}

In this section we will reformulate the BS equation Eq.~(\ref{BSE})
as an integral equation in terms of the weight functions.  This is the
central result of our work.
For notational convenience we first rewrite the scattering kernel
$I(p,q;P)$ as follows
\begin{eqnarray}
  & &  I(p,q;P)\\
  & = &\sum_{\rm ch}\int\limits_{0}^{\infty }
                d\gamma\int\limits_{\Omega}
		d\vec\xi
        	\frac{\rho_{\rm ch}(\gamma,\vec\xi)}
		{\gamma-\left(
			a_{\rm ch}\,q^2+b_{\rm ch}\,
                                p\cdot q+c_{\rm ch}\,p^2+d_{\rm ch}\,P^2
				+e_{\rm ch}\,q\cdot P+f_{\rm ch}\,p\cdot P
			\right)-i\epsilon
		}\; ,\nonumber
    \label{krnl_PTIR2}
\end{eqnarray}
where $\{a_{\rm ch},b_{\rm ch},c_{\rm ch},\dots,f_{\rm ch}\}$ are different
linear combinations of $\xi_i$ in each of the three channels,
{\it ch}$=\{st\},\{tu\},\{us\}$.  Substituting the integral representation
of the partial wave BS amplitude Eq.~(\ref{ptlw_BS_PTIR}),
together with the PTIR for the scattering kernel
Eq.~(\ref{krnl_PTIR2}), into the BS equation Eq.~(\ref{BSE}) and
after multiplying the
propagators $D(P/2+p)D(P/2-p)$ into both sides,
one can perform the $q$-integral without relying on Wick rotation
imposing the condition
for the dummy parameter $n$ such that $n+1 > l/2$.  This ensures
that the $q$-integral is finite.
%Due to the self-reproducing
%property of the solid harmonics this integral is again proportional
%to the solid harmonics ${\cal Y}_\ell^{\ell_z}(\Lambda^{-1}(P) p)$.
%Multiplying the propagators $D(P/2+p)D(P/2-p)$ into both sides
%and absorbing them into the RHS expression using Feynman parameterization,
%the BS equation then becomes
%\begin{eqnarray}
%	\Phi^{[\ell,\ell_z]}(p,P) & = &
%	-i{\cal Y}_\ell^{\ell_z}(\Lambda^{-1}(P) p)
%	\int\limits_{-\infty}^{\infty}d
%        \bar\alpha\int\limits_{-1}^{1}d\bar z,
%	\frac{1}
%	{\left[m^2+\bar\alpha
%			-\left(p^2 + \bar z p\cdot P + P^2/4\right)
%			-i\epsilon\right]^{n+2}
%	}
%	\nonumber\\
%	& & \qquad\times
%	\sum_{\rm ch}\int\limits_{0}^{\infty } d\gamma\int\limits_{\Omega}
%	\, \rho_{\rm ch}(\gamma,\vec\xi) \,
%	\int\limits_{-\infty}^{\infty}d\alpha\int\limits_{-1}^{1}d
%        z \,
%		K^{[\ell]}_n(\bar\alpha,\bar z;\alpha,z)
%		\,\, \varphi^{[\ell]}_n(\alpha,z)\;.
%	\label{BSE2}
%\end{eqnarray}
Using the uniqueness theorem of PTIR\cite{Nakanishi_graph},
we obtain the following integral equation for $\varphi^{[\ell]}_n(\alpha,z)$:
\begin{equation}
	\varphi^{[\ell]}_n(\bar\alpha,\bar z)=
	  \int\limits_{-\infty}^{\infty}d\alpha
	\int\limits_{-1}^{1}dz \,
	{\cal K}^{[\ell]}_n(\bar\alpha,\bar z;\alpha,z)
	\varphi^{[\ell]}_n(\alpha,z)\;.
	\label{eqn_wght}
\end{equation}
We have defined the total kernel function
${\cal K}^{[\ell]}_n(\bar\alpha,\bar z;\alpha,z)$, which is a superposition
of the kernel functions $K^{[\ell]}_n(\bar\alpha,\bar z;\alpha,z)$,
which in turn are defined with a fixed
parameter set of the scattering kernel
$\{a_{\rm ch},b_{\rm ch},c_{\rm ch},\dots,f_{\rm ch}\}$,
\begin{equation}
	{\cal K}^{[\ell]}_n(\bar\alpha,\bar z;\alpha,z)\equiv
	  \sum_{\rm ch}\int\limits_{0}^{\infty } d\gamma\int\limits_{\Omega}
           d\vec\xi
	  \, \rho_{\rm ch}(\gamma,\vec\xi) \,
	  K^{[\ell]}_n(\bar\alpha,\bar z;\alpha,z).
	\label{totl_krnl}
\end{equation}
The kernel function with a fixed kernel parameter set is
\begin{eqnarray}
	K_n^{[\ell]}(\bar\alpha,\bar z;\alpha,z)&=&\frac{1}{(4\pi)^2}
		\frac{1}{2} \left(-{b\over 2}\right)^l\int_{0}^{\infty}dy
		{\rm Pf}\,\cdot \frac{y^{n+1} (a+y)^{n-1-l}}
		{\left[A(\alpha,z)y^2+B(\alpha,z)y+C\right]^{n+1}}
	\label{gnrl_n_kernel}\\
	& & \, \times \frac{\partial}{\partial \bar\alpha}\bar\alpha^n
	  \left[\theta(\bar\alpha)-\theta\left(\bar\alpha-R(\bar z, G(z;y))
	  \frac{A(\alpha,z)y^2+B(\alpha,z)y+C}{c y+\Delta}\right)\right],
	\nonumber
\end{eqnarray}
with functions $A(\alpha,z)$, $B(\alpha,z)$, $R(\bar z,z)$ and $G(z;y)$ as
\begin{eqnarray}
	& &A(\alpha,z)  =  \alpha + m^2 -(1-z^2){P^2\over 4},
	\nonumber \\
	& &B(\alpha,z)  = a\alpha+\gamma + (a-c)\left(m^2-{P^2\over 4}\right)
				-(4d-2ez){P^2\over 4},
	\nonumber \\
	& &G(z;y)  =  \frac{(f-b/2 z)y+af-eb/2}{c y+\Delta}\;,
	\label{para_def} \\
	\nonumber \\
	& &R(\bar z,z) = \frac{1-\bar z}{1- z}\theta(\bar z - z) +
	                  \frac{1+\bar z}{1+ z}\theta(z - \bar z ),
	\nonumber
\end{eqnarray}
and the constants; $C  =  a \gamma - \Delta\left(m^2-{P^2\over 4}\right)
-(4ad-e^2){P^2\over 4}$, $\Delta =ac-{b^2\over 4}$.
%\begin{eqnarray}
%	& &C  =  a \gamma - \Delta\left(m^2-{P^2\over 4}\right)
%			-(4ad-e^2){P^2\over 4}\;,  \\
%	& &\Delta =ac-{b^2\over 4}\;.	\nonumber
%\end{eqnarray}
The integration over the Feynman parameter $y$ should be carried out
using the Hadamard finite part prescription:
\begin{equation}
	{\rm Pf}\,\cdot {1 \over x^n} =
	\lim_{\epsilon\rightarrow 0} {\rm Re}{1 \over (x\pm i\epsilon)^n}\;,
	\quad \hbox{for integer }n.
\end{equation}
This is consistent with the ordinary $i\epsilon$ prescription
for a calculation of Feynman diagrams in momentum space
\cite{Nakanishi_graph} and we keep $\epsilon$ finite but small
for numerical studies to regularize possible singularities of the kernel.
A detailed analysis of singularities of the kernel function may be
found in Ref.~6.
%\cite{KandW}.

Since the weight functions $\rho_{\rm ch}(\gamma,\vec\xi)$
for the scattering kernel are real functions by their construction,
the total kernel function ${\cal K}^{[\ell]}_n(\bar\alpha,\bar z;\alpha,z)$
is real, so that
the Eq.~(\ref{eqn_wght}) is a real integral equation with
two variables $\alpha$ and $z$.  Thus we have transformed the
BS equation, which is a singular integral equation of
complex distributions, into a real integral equation which is
frame-independent.
As is evident from Eq.~(\ref{rln_wghtfnc})
if $n$ is chosen sufficiently large these
can be always transformed into arbitrarily smooth functions suitable for
numerical studies.

\section{Numerical Studies}
\label{Results}

To illustrate the approach we have numerically solved the following
two explicit examples : \\
	(a) Scalar-scalar ladder model with massive scalar exchange:
	The simple $t$-channel one-$\sigma$-exchange kernel is given by
	\begin{equation}
		I(p,q;P)=\frac{g^2}{m_\sigma^2-(p-q)^2-i\epsilon}
		\label{pure_ladder_kernel}
	\end{equation}
	The BS equation with this kernel together with perturbative
	 propagator $D^0$ is often referred to as the
	``scalar-scalar ladder model''\cite{Nakanishi_survey}.
	The Wick rotated BS equation for this kernel has been studied
	numerically\cite{L+M}.  We use this kernel as a check of our
	calculations.\\
	(b) Generalized ladder kernel:
	A sum of the one-$\sigma$-exchange kernel
	Eq.~(\ref{pure_ladder_kernel})
	and a generalized kernel terms in the $st$-channel
	with constant kernel parameter sets
	$\{\vec\xi^{st}\}$ weighted with a factor $(1/4)[g^2/(4\pi)^2]$.
	We generated $\{\vec\xi^{st}\}$ by a random number generator.
	After the Wick rotation this kernel
	becomes complex due to the $p\cdot P$ and $q\cdot P$ terms, so that
	solving the BS amplitude as a function of Euclidean relative momentum
	would be difficult in this case.

The scattering kernel (a), i.e., the one-$\sigma$-exchange
kernel corresponds to choosing for the kernel in Eq.~(\ref{krnl_PTIR}) say:
$\rho_{tu}=\rho_{us}=0$ and in the $st$-channel $\gamma=m_\sigma^2$,
and $a_{st}=c_{st}=1$, $b_{st}=-2$, $d_{st}=e_{st}=f_{st}=0$ with
$\rho_{st}$ to be some appropriate product of
$\delta$-functions multiplied by $g^2$.
It is often convenient (and traditional) to factorize out
the coupling constant $g^2$ together with a factor of $(4\pi)^2$, and
define the ``eigenvalue'' $\lambda=g^2/(4\pi)^2$.
Then we solve the equation as a eigenvalue
problem with a fixed bound state mass $P^2$.
To regularize singularities of kernel function
we kept the regularization parameter
$\epsilon$ in the Hadamard finite part Pf$\cdot 1/x^n$
small but finite, typically $\sim 10^{-5}$ or less.
We began by discretizing the $\alpha$ and $z$ axes and then solved
the integral equation by iteration from some initial
assumed weight function.  We verified that our solutions were independent of
$\epsilon$ provided it was chosen sufficiently small.
It was also, of course, confirmed that the solution
was robustly independent of the choice of
starting weight function, the number
of grid points, and the maximum grid value of $\alpha$ (when the latter
two of these were chosen suitably large).
We found that the convergence and the stability of the eigenvalues
with varying the number of grid points, $\epsilon$, and $\alpha_{\rm max}$
are excellent.
We have solved over a range of $P^2$ and $m_\sigma$ and all solutions
reproduced the eigenvalues obtained in Euclidean space
after a Wick rotation by Linden and Mitter \cite{L+M}.
For a nontrivial example of the scattering kernel, we studied the model
kernel (b) symmetrized in $\bar z \rightarrow -\bar z$ so that we obtain
appropriate symmetry for $s$-wave normal solutions.
We found that the convergence and the stability of the eigenvalues
with varying the number of grid points, $\epsilon$, and $\alpha_{\rm max}$
are excellent as for the pure ladder case \cite{KandW}.  Additional
details and figures can be found elsewhere \cite{KandW}.

\section{Summary}
\label{Conclusions}

We have derived a real integral equation for the weight function of the
scalar-scalar Bethe-Salpeter (BS) amplitude from the BS equation for
scalar theories without derivative coupling.
This was achieved using the perturbation theory integral representation
(PTIR), which is an extension of the spectral representation for two-point
Green's functions, for both the scattering kernel [Eq.~(\ref{krnl_PTIR})]
and the BS amplitude itself [Eqs.~(\ref{ptlw_BS_PTIR})].
The uniqueness theorem of the PTIR and the appropriate application
of Feynman parameterization then led to the central result of the paper
given in Eq.~(\ref{eqn_wght}).
We demonstrated that this integral equation is numerically tractable
for both the pure ladder case and an arbitrary generalization of this.
We have verified that our numerical solutions agree with those previously
obtained from a Euclidean treatment of the pure ladder limit.

This represents a potentially powerful new approach to obtaining solutions of
the BS equation and additional results and applications are currently being
investigated.  The separable kernel case should be studied,
since it has exact solutions and so is a further independent
test of the approach
developed here.  While our detailed discussions were limited to the equal mass
case, it is worthwhile to investigate generalizations including
nonperturbative constituent propagators and the heavy-light bound state
limit to see under which conditions an approximate Klein-Gordon equation
can result.
It is also important to find a means to generalize
this approach to include derivative couplings and fermions.

\section{Acknowledgements}

This work was supported by the Australian Research Council and
also in part by grants of
supercomputer time from the U.S. National Energy Research Supercomputer
Center and the Australian National University
Supercomputer Facility.

\section{References}
%=======================================================================
%          Bibliography:
%-----------------------------------------------------------------------
%%%%%%%%%%%%%%%%%

%%%%%%%%%%%%%%%%%

\end{document}